\documentclass[
  reprint, 
  superscriptaddress,
  amsmath, amssymb,
  aps,
  prl,
  bibnotes,
]{revtex4-2}
\usepackage{graphicx}
\usepackage{dcolumn}
\usepackage{bm}
\usepackage{braket}
\usepackage{booktabs}
\usepackage{tikz}
\usetikzlibrary{arrows.meta,positioning,calc,fit,backgrounds}
\newcommand{\prsec}[1]{\noindent\textit{#1.---}}
\usepackage{hyperref}

\begin{document}

\title{\textbf{Physics-inspired transformer quantum states via latent imaginary-time evolution}}

\author{Kimihiro Yamazaki}
\email{Contact: k-yamazaki@ist.osaka-u.ac.jp}
\affiliation{Graduate School of Information Science and Technology, The University of Osaka, 1-5 Yamadaoka, Suita, Osaka, Japan.}
\affiliation{Fujitsu Laboratories, Fujitsu Limited, 4-1-1 Kamikodanaka, Nakahara-ku, Kawasaki, Kanagawa, Japan}

\author{Itsushi Sakata}
\affiliation{Center for Advanced Intelligence Project, RIKEN, 1-4-1 Nihonbashi, Chuo-ku, Tokyo, Japan.}

\author{Takuya Konishi}
\affiliation{Graduate School of Information Science and Technology, The University of Osaka, 1-5 Yamadaoka, Suita, Osaka, Japan.}
\affiliation{Center for Advanced Intelligence Project, RIKEN, 1-4-1 Nihonbashi, Chuo-ku, Tokyo, Japan.}

\author{Yoshinobu Kawahara}
\affiliation{Graduate School of Information Science and Technology, The University of Osaka, 1-5 Yamadaoka, Suita, Osaka, Japan.}
\affiliation{Center for Advanced Intelligence Project, RIKEN, 1-4-1 Nihonbashi, Chuo-ku, Tokyo, Japan.}

\date{\today}

\begin{abstract}

Neural quantum states (NQS) are powerful ans\"atze in the variational Monte Carlo framework, yet their architectures are often treated as black boxes. 
We propose a physically transparent framework in which NQS are treated as neural approximations to latent imaginary-time evolution.
This viewpoint suggests that standard Transformer-based NQS (TQS) architectures correspond to physically unmotivated effective Hamiltonians dependent on imaginary time in a latent space.
Building on this interpretation, we introduce physics-inspired transformer quantum states (PITQS), which enforce a static effective Hamiltonian by sharing weights across layers and improve propagation accuracy via Trotter--Suzuki decompositions without increasing the number of variational parameters.
For the frustrated $J_1$-$J_2$ Heisenberg model, our ans\"atze achieve accuracies comparable to or exceeding state-of-the-art TQS while using substantially fewer variational parameters. 
This study demonstrates that reinterpreting the deep network structure as a latent cooling process enables a more physically grounded, systematic, and compact design, thereby bridging the gap between black-box expressivity and physically transparent construction.

\end{abstract}

\maketitle

\prsec{Introduction}Accurately obtaining ground-state wave functions of quantum many-body systems, especially strongly correlated systems~\cite{imada1998rev}, is a grand challenge in modern physics because of the complexity of many-body correlations.
In recent years, neural quantum states (NQS)~\cite{carleo2017solving} have emerged as a powerful class of variational ans\"atze within variational Monte Carlo (VMC)~\cite{mcmillan1965ground}.
In particular, Transformer-based NQS (TQS)~\cite{viteritti2023transformer0,PhysRevB.111.134411,qxc3-bkc7,gu2025solvinghubbardmodelneural} have achieved state-of-the-art accuracy on some of the most challenging benchmark problems, notably the frustrated $J_1$-$J_2$ Heisenberg model~\cite{rende2024simple}.

Despite these successes, existing NQS architectures, particularly TQS~\cite{viteritti2023transformer0,PhysRevB.111.134411,qxc3-bkc7,gu2025solvinghubbardmodelneural}, largely repurpose generic machine-learning designs and thus remain black-box ans\"atze.
As a result, improved accuracy is often achieved through substantial overparameterization, while providing limited physical insight into how and why the ansatz attains its performance.
This black-box nature also makes systematic architectural refinement difficult, since there is no unified principle linking architectural changes to controlled improvements in accuracy.

In parallel to the development of variational ans\"atze, a long-standing and physically transparent route to ground states is provided by \emph{imaginary-time evolution (ITE)}.
The fundamental principle of ITE relies on cooling an arbitrary initial state $|\Psi_0\rangle$ toward the ground state $|\Psi\rangle$ as
\begin{equation}
|\Psi\rangle \propto e^{-\beta \hat{\mathcal{H}}} |\Psi_0\rangle,
\label{eq:ite}
\end{equation}
where $\hat{\mathcal{H}}$ is the Hamiltonian and $\beta$ is a sufficiently large imaginary time~\cite{PhysRev.84.350}.
The auxiliary-field quantum Monte Carlo (AF-QMC) method~\cite{PhysRevD.24.2278,SUGIYAMA19861} realizes ITE by employing the Hubbard--Stratonovich transformation.
Although this approach yields exact estimates for specific Hamiltonians, the explicit stochastic sampling suffers from a severe sign problem, particularly in frustrated spin and fermion systems, which strictly limits its applicability.
Even recent deep Boltzmann machine constructions of exact ITE~\cite{carleo2018constructing} essentially inherit the sign problem in frustrated systems. 
Thus, a robust, sign-problem-free integration of NQS and ITE for generic frustrated settings remains a significant challenge.

In this Letter, we bridge the gap between black-box NQS and physically transparent construction by introducing a framework based on \emph{latent imaginary-time evolution (LITE)}.
We represent the NQS ansatz through a cooling process within a latent space, inspired by the imaginary-time picture in Eq.~(\ref{eq:ite}) within the VMC scheme.
In this scheme, instead of directly using the bare physical Hamiltonian $\hat{\mathcal{H}}$, we model the LITE with a learnable effective Hamiltonian in the latent space, which is parameterized by a neural network.
Crucially, this variational formulation makes our approach inherently free from the sign problem.

From the perspective of LITE, it becomes clear that existing TQS architectures~\cite{viteritti2023transformer0,PhysRevB.111.134411,qxc3-bkc7,gu2025solvinghubbardmodelneural} effectively imply that the cooling process is driven by an effective Hamiltonian with imaginary-time dependence. 
This variation at every evolution step represents an unphysical redundancy for ground-state cooling and leads to significant overparameterization. To address this, we propose \emph{physics-inspired transformer quantum states (PITQS)}. Since the cooling process is naturally governed by a single Hamiltonian independent of imaginary time, we enforce a static effective Hamiltonian by sharing weights across all layers.

This framework provides immediate physical insight into the network architecture and highlights a fundamental limitation of existing TQS. In particular, once each layer of the TQS is identified as an approximation to the latent imaginary-time propagator generated by non-commuting terms in the effective Hamiltonian, the standard TQS update is naturally interpreted as a first-order Lie--Trotter decomposition~\cite{lie1888theorie,trotter_original}. 
As in any first-order decomposition, finite per-layer imaginary-time steps induce systematic Trotter errors. Motivated by this observation, we systematically implement higher-order Trotter--Suzuki decompositions~\cite{doi:10.1137/0705041,SUZUKI1990319,BLANES2002313} of the static effective Hamiltonian, thereby improving the propagator accuracy without increasing the number of variational parameters.

We demonstrate the capabilities of our framework on the frustrated $J_1$-$J_2$ Heisenberg model, which serves as an established benchmark in the NQS domain~\cite{chen2024empowering,rende2024simple}.
Our results show that a static effective Hamiltonian is sufficiently expressive to capture intricate quantum correlations, indicating that much of the parameter budget in conventional TQS, which implicitly employ effective Hamiltonians with imaginary-time dependence, is redundant.
Guided by this physical perspective, we explicitly design both the effective Hamiltonian and the Trotter--Suzuki scheme, achieving accuracies that are comparable to, and in some cases exceed, state-of-the-art benchmarks while using substantially fewer variational parameters.
Overall, our approach elevates the design of the NQS architecture from an empirical heuristic to a systematic, physically grounded construction.

\begin{figure}[t!]
    \centering
    \includegraphics[width=\linewidth]{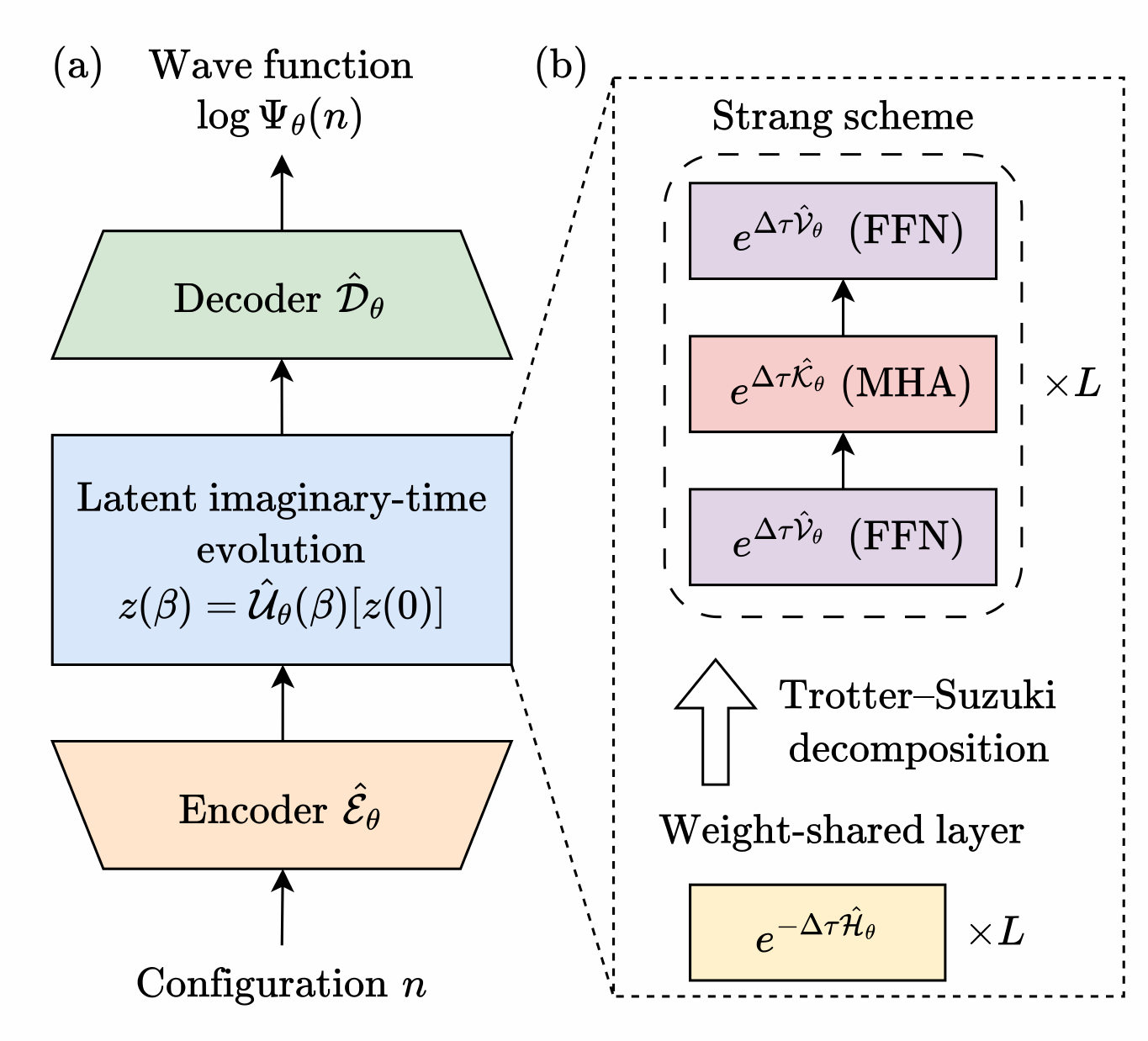}
    \caption{PITQS overview.
    (a) A configuration $n$ is mapped by the encoder $\hat{\mathcal{E}}_\theta$ to initial latent states $z(0)$, evolved by the LITE operator $\hat{\mathcal{U}}_\theta(\beta)$ to $z(\beta)$, and decoded by $\hat{\mathcal{D}}_\theta$ to the wave-function log-amplitude $\log\Psi_\theta(n)$.
    (b) In PITQS, $\hat{\mathcal{U}}_\theta(\beta)$ is implemented as $L$ weight-shared layers approximating $e^{-\Delta\tau\hat{\mathcal{H}}_\theta}$ with a single effective Hamiltonian $\hat{\mathcal{H}}_\theta$; each step uses a Trotter--Suzuki decomposition (Strang decomposition scheme shown) that alternates the non-local operator $\hat{\mathcal{K}}_\theta$ and the local operator $\hat{\mathcal{V}}_\theta$.}
    \label{fig:overview}
\end{figure}

\prsec{Latent imaginary-time evolution}Here we formulate a framework to construct an NQS ansatz via LITE and optimize it within the VMC scheme.
Working in a discrete configuration basis $\{|n\rangle\}$ (for an $N$ spin-$1/2$ system, $n$ labels spin configurations of the Hilbert space), we parameterize $\Psi_\theta(n)=\langle n|\Psi_\theta\rangle$
by multiple deep neural networks with variational parameters $\theta$, composed of three logical blocks:
a latent encoder $\hat{\mathcal{E}}_\theta$, a LITE operator $\hat{\mathcal{U}}_\theta(\beta)$, and a wave-function decoder $\hat{\mathcal{D}}_\theta$.
The total variational ansatz is defined as the composition
\begin{equation}
  \log\Psi_{\theta}(n)
  \;=\;
  \hat{\mathcal{D}}_\theta\Bigl[ \hat{\mathcal{U}}_\theta(\beta)\bigl[\hat{\mathcal{E}}_\theta[n]\bigr]\Bigr]\in\mathbb{C},
\end{equation}
where $\beta$ denotes the total imaginary-time in the latent space (Fig.~\ref{fig:overview}(a)).

First, the latent encoder operator $\hat{\mathcal{E}}_\theta$ maps a discrete configuration $n$ to a latent state
$z(0)\in\mathbb{R}^{N_{\mathrm{tok}}\times d}$.
Here $z(0)$ is represented by $N_{\mathrm{tok}}$ latent tokens $z_i(0)\in\mathbb{R}^d$, where $i$ is the latent-token index and $d$ is the token dimension.
Introducing latent tokens is analogous in spirit to introducing auxiliary fields in AF-QMC~\cite{PhysRevD.24.2278,SUGIYAMA19861}; it provides extra degrees of freedom through which nontrivial correlations can be represented efficiently. However, unlike AF-QMC, our latent variables are learnable and optimized end-to-end within VMC, rather than being stochastically sampled and integrated out. This latent-space viewpoint thus provides a physically motivated way to enlarge the variational manifold while retaining the sign-problem-free nature of the VMC formulation.

Next, $\hat{\mathcal{U}}_\theta(\beta)$ performs the LITE cooling over a total imaginary time $\beta$ and realizes a low-energy projection in the latent space.
In our approach, the LITE operator $\hat{\mathcal{U}}_\theta(\beta)$ is implemented by a neural network, and we write
\begin{equation}
  z(\beta)=\hat{\mathcal{U}}_\theta(\beta)\bigl[z(0)\bigr].
\end{equation}
It is convenient to interpret this action in terms of a latent state $z(\tau)\in\mathbb{R}^{N_{\mathrm{tok}}\times d}$, $0\le\tau\le\beta$, with $z(0)=\hat{\mathcal{E}}_\theta[n]$ and $z(\beta)=\hat{\mathcal{U}}_\theta(\beta)[z(0)]$.
In the continuum limit, one may regard $z(\tau)$ as obeying the latent imaginary-time evolution equation $d z(\tau)/d\tau= -\,\hat{\mathcal{H}}_\theta\, [z(\tau)]$, whose solution yields
\begin{equation}
  z(\beta)=e^{-\beta\hat{\mathcal{H}}_\theta}[z(0)].
  \label{eq:latent_ite_continuous}
\end{equation}
Here, $\hat{\mathcal{H}}_\theta$ denotes an effective Hamiltonian parameterized by a neural network in the latent space.
This is directly analogous to the ITE principle in Eq.~(\ref{eq:ite}), except that the latent state evolves under the effective Hamiltonian $\hat{\mathcal{H}}_\theta$ rather than the physical Hamiltonian $\hat{\mathcal{H}}$.

Finally, the wave-function decoder $\hat{\mathcal{D}}_\theta$ maps the evolved latent state $z(\beta)$ to the wave-function log-amplitude $\log\Psi_\theta(n)$.

\prsec{Relation to existing TQS}We re-examine the existing TQS architecture~\cite{viteritti2023transformer0,rende2024simple,PhysRevB.111.134411,qxc3-bkc7,gu2025solvinghubbardmodelneural}
through the lens of LITE.
In this viewpoint, existing TQS can be organized into an embedding that maps physical degrees of freedom to an initial latent representation, a wave-function decoder that maps the final latent representation to the variational many-body wave function, and a depthwise latent update implemented by repeated multi-head attention (MHA) and feed-forward network (FFN) transformations acting on latent tokens.

First, we identify our latent encoder and decoder.
The embedding used in standard TQS~\cite{viteritti2023transformer0,rende2024simple,PhysRevB.111.134411} corresponds to the latent encoder $\hat{\mathcal{E}}_\theta$: it partitions each configuration $n$ into local patches and maps them to $z(0)\in\mathbb{R}^{N_{\mathrm{tok}}\times d}$.
For a two-dimensional $M\times M$ lattice with square patches of size $b\times b$, this yields $N_{\mathrm{tok}}=(M/b)^2$ tokens.
On the decoder side, the standard TQS formulation~\cite{viteritti2023transformer0,rende2024simple,PhysRevB.111.134411} is as follows:
\begin{equation}
  \hat{\mathcal{D}}_\theta[z(\beta)]
  =\sum_{a=1}^{d}\log\cosh\,\!\bigl(w_a^{\top}\zeta+b_a\bigr),
  \label{eq:tqs_decoder_logcosh}
\end{equation}
where $\zeta=\sum_{i=1}^{N_{\mathrm{tok}}}z_{i}(\beta)\in\mathbb{R}^{d}$ and $\{w_{a},b_{a}\}_{a=1}^{d}$ are complex-valued variational parameters.

Next, focusing on the stacked TQS layers, we can interpret the updates driven by MHA and FFN as the LITE.
We discretize Eq.~(\ref{eq:latent_ite_continuous}) by partitioning $\tau\in[0,\beta]$ into $L$ slices with $\Delta\tau=\beta/L$, introducing $\tau_\ell=\ell\Delta\tau$, and defining the discrete latent state
$z^{(\ell)} \equiv z(\tau_\ell)$ $(\ell=0,\dots,L)$, so that $z^{(0)}=z(0)$ and $z^{(L)}=z(\beta)$.
Within this notation, the $\ell$-th layer updates the latent state from $z^{(\ell-1)}$ to $z^{(\ell)}$ and can be viewed as a neural approximation to a short imaginary-time propagator acting in the latent space.

In General, a quantum many-body Hamiltonian $\hat{\mathcal{H}}$ can be written as the sum of a non-local term $\hat{\mathcal{K}}$ and a local (on-site) term $\hat{\mathcal{V}}$.
Motivated by this structure, we interpret the two sub-maps inside a TQS layer as providing an analogous decomposition in the latent space: the MHA sublayer updates each latent token through an aggregation over all other tokens, thereby enabling long-range token–token couplings and playing the role of an effective non-local interaction term, whereas the FFN sublayer applies the same nonlinear map independently to each token and thus plays the role of an effective on-site term.
Concretely, denoting the MHA and FFN residual updates in layer $\ell$ by the maps
$\hat{\mathcal{K}}^{(\ell)}_{\theta}$ and $\hat{\mathcal{V}}^{(\ell)}_{\theta}$, respectively, we define the layer-$\ell$ effective Hamiltonian in the latent space as
\begin{equation}
  \hat{\mathcal{H}}^{(\ell)}_{\theta} \equiv -\Bigl(\hat{\mathcal{V}}^{(\ell)}_{\theta}+\hat{\mathcal{K}}^{(\ell)}_{\theta}\Bigr),
\end{equation}
where the minus sign is chosen so that the ITE cooling corresponds to applying the LITE operator.
With this definition, $\hat{\mathcal{H}}^{(\ell)}_{\theta}$ has the typical structure of a quantum many-body Hamiltonian as a sum of a local term and a non-local term, making it clear that the TQS naturally induces an effective Hamiltonian in the latent space in a form well suited to representing quantum many-body systems.

Each layer of the existing TQS applies MHA and FFN sequentially, which is naturally interpreted as a first-order Lie--Trotter decomposition~\cite{lie1888theorie,trotter_original}
of the short imaginary-time propagator; for a single step at layer $\ell$,
\begin{equation}
e^{-\Delta\tau \hat{\mathcal{H}}^{(\ell)}_{\theta}}
=
e^{\Delta\tau(\hat{\mathcal{V}}^{(\ell)}_{\theta}+\hat{\mathcal{K}}^{(\ell)}_{\theta})}
\;\approx\;
e^{\Delta\tau\hat{\mathcal{V}}^{(\ell)}_{\theta}}\,
e^{\Delta\tau\hat{\mathcal{K}}^{(\ell)}_{\theta}}.
\label{eq:tqs_lie_trotter}
\end{equation}
Because $[\hat{\mathcal{K}}^{(\ell)}_{\theta},\hat{\mathcal{V}}^{(\ell)}_{\theta}]\neq 0$, this decomposition incurs Trotter errors scaling locally as $\mathcal{O}(\Delta\tau^{2})$.
Approximating each component by the Euler-type update yields the concrete layer update
\begin{equation}
\begin{aligned}
  \tilde{z}^{(\ell)} &= e^{\Delta\tau\hat{\mathcal{K}}^{(\ell)}_{\theta}}[z^{(\ell-1)}]
  \approx z^{(\ell-1)} + \Delta\tau\,\hat{\mathcal{K}}^{(\ell)}_{\theta}[z^{(\ell-1)}], \\
  z^{(\ell)} &= e^{\Delta\tau\hat{\mathcal{V}}^{(\ell)}_{\theta}}[\tilde{z}^{(\ell)}]
  \approx \tilde{z}^{(\ell)} + \Delta\tau\,\hat{\mathcal{V}}^{(\ell)}_{\theta}[\tilde{z}^{(\ell)}],
\end{aligned}
\end{equation}
where $\tilde z^{(\ell)}$ denotes an intermediate latent state after applying the non-local update within the $\ell$-th layer. This matches the standard TQS layer ordering (MHA $\rightarrow$ FFN) within each layer.
Finally, stacking $L$ layers implements the LITE operator as an ordered product of short imaginary-time propagators,
\begin{equation}
  \hat{\mathcal{U}}_{\theta}(\beta)
  \;\approx\;
  T_\tau\prod_{\ell=1}^{L}
  e^{\Delta\tau\hat{\mathcal{V}}^{(\ell)}_{\theta}}\,
  e^{\Delta\tau\hat{\mathcal{K}}^{(\ell)}_{\theta}},
  \label{eq:time_ordered_product}
\end{equation}
where $T_\tau$ denotes the imaginary-time-ordered product in the latent space.

Importantly, in conventional TQS architectures~\cite{viteritti2023transformer0,rende2024simple,PhysRevB.111.134411}, the weights are not shared across layers, so the induced latent-space operators vary with $\ell$:
for $\ell\neq \ell'$, one generally has $\hat{\mathcal{H}}^{(\ell)}_{\theta}\neq \hat{\mathcal{H}}^{(\ell')}_{\theta}$.
Equivalently, the standard depth-$L$ TQS realizes effective Hamiltonians dependent on latent imaginary time.
However, from a physical standpoint, the VMC target problem is specified by a single, imaginary-time independent Hamiltonian $\hat{\mathcal{H}}$, and ITE cooling is an autonomous process.
Therefore, once the stacked TQS layers are interpreted as implementing LITE, the layer-wise variation of $\hat{\mathcal{H}}^{(\ell)}_{\theta}$ constitutes a physically redundant freedom, leading to substantial overparameterization: the number of variational parameters scales linearly with $L$
even though the underlying mechanism is governed by a single Hamiltonian.

\prsec{Physics-inspired transformer quantum states}Guided by this viewpoint, we introduce PITQS, a physics-inspired construction of $\hat{\mathcal{U}}_\theta(\beta)$ that enforces a single effective Hamiltonian independent of imaginary time and systematically improves the propagation accuracy by higher-order Trotter--Suzuki decompositions (Fig.~\ref{fig:overview}(b)).
Concretely, instead of the layer-dependent effective Hamiltonians $\{\hat{\mathcal{H}}_{\theta}^{(\ell)}\}_{\ell=1}^{L}$ implicit in standard TQS, we impose a single weight-shared effective Hamiltonian
\begin{equation}
\hat{\mathcal{H}}_\theta = -\bigl(\hat{\mathcal{V}}_\theta+\hat{\mathcal{K}}_\theta\bigr),
\end{equation}
and approximate the LITE operator,
\begin{equation}
  \hat{\mathcal{U}}_\theta(\beta)
  \;=\;
  \bigl(e^{-\Delta\tau\,\hat{\mathcal{H}}_\theta}\bigr)^{L}=e^{-\beta\hat{\mathcal{H}}_\theta}.
  \label{eq:pitqs_U_static}
\end{equation}
This corresponds precisely to realizing the ITE of Eq.~(\ref{eq:ite}) in the latent space.
Crucially, the weight-sharing constraint across layers removes the physically redundant layer-wise variation present in standard TQS [Eq.~(\ref{eq:time_ordered_product})] and reduces the number of variational parameters by a factor of order $L$.

Furthermore, the LITE viewpoint suggests a systematic route beyond the first-order Lie--Trotter decomposition.
We replace the single-step propagator with an $m$-th Trotter--Suzuki decomposition,
\begin{equation}
e^{-\Delta\tau \hat{\mathcal{H}}_\theta}
\approx
\prod_{i=1}^{k}
\exp\!\left(a_{i}^{(m)}\Delta\tau\,\hat{\mathcal{V}}_\theta\right)\,
\exp\!\left(b_{i}^{(m)}\Delta\tau\,\hat{\mathcal{K}}_\theta\right),
\label{eq:pitqs_suzuki_trotter}
\end{equation}
where $k$ is the number of stages and $\{a_{i}^{(m)}, b_{i}^{(m)}\}$ are fixed coefficients defining an $m$-th order decomposition with local Trotter error $\mathcal{O}(\Delta\tau^{m+1})$ and satisfying
$\sum_i a_{i}^{(m)}=\sum_i b_{i}^{(m)}=1$.
For example, the second-order Strang ($m=2$) decomposition~\cite{doi:10.1137/0705041} takes the form
\begin{equation}
  e^{-\Delta\tau \hat{\mathcal{H}}_\theta}
  \;\approx\;
  e^{\frac{\Delta\tau}{2}\hat{\mathcal{V}}_\theta}\,
  e^{\Delta\tau\hat{\mathcal{K}}_\theta}\,
  e^{\frac{\Delta\tau}{2}\hat{\mathcal{V}}_\theta},
  \label{eq:pitqs_strang}
\end{equation}
with local error $\mathcal{O}(\Delta\tau^{3})$.
Furthermore, in our experiments, we employ two fourth-order schemes, due to Suzuki~\cite{SUZUKI1990319} and Blanes--Moan~\cite{BLANES2002313}.
Importantly, increasing the decomposition order only changes these fixed coefficients and introduces no additional variational parameters, thereby improving the accuracy of the latent imaginary-time propagation without enlarging the parameter budget.

\begin{table}[t!]
  \centering
  \caption{
    Energies per site $E$ obtained with the PITQS and standard TQS using the Lie--Trotter scheme. Parentheses indicate statistical (Monte Carlo) errors. We also report the corresponding total imaginary time $\beta$ and the number of variational parameters $N_p$ for each setting.
  }
  \begin{ruledtabular}
    \begin{tabular}{ccccc}
      $\beta$ & $E$\,(TQS) & $N_{p}$\,(TQS) & $E$\,(PITQS) & $N_{p}$\,(PITQS) \\
      \hline
      1.0 & $-0.49593(11)$ & 81,800  & $-0.49596(14)$ & 44,890 \\
      2.0 & $-0.49652(9)$  & 155,620 & $-0.49669(9)$  & 44,890 \\
      3.0 & $-0.49656(11)$ & 229,440 & $-0.49676(10)$  & 44,890 \\
      4.0 & $-0.49651(10)$ & 303,260 & $-0.49696(9)$  & 44,890 \\
    \end{tabular}
    \end{ruledtabular}
  \label{tab:result1}
\end{table}

\begin{table}[t]
  \centering
  \caption{
  Energies per site at $\beta=2.0$ for several PITQS schemes and standard TQS baselines ($\beta=0.5$ and $2.0$). $N_p$ denotes the number of variational parameters.
  }
  \begin{ruledtabular}
  \begin{tabular}{lcc}
    Scheme & Energy per site & $N_p$ \\
    \hline
    TQS ($\beta=0.5$) & $-0.49318(15)$ & 44{,}890 \\
    TQS ($\beta=2.0$)& $-0.49652(9)$  & 155{,}620 \\
    PITQS (Lie--Trotter)  & $-0.49669(9)$  & 44{,}890 \\
    PITQS (Strang)        & $-0.49671(9)$  & 44{,}890 \\
    PITQS (Suzuki)        & $-0.49697(9)$  & 44{,}890 \\
    PITQS (Blanes--Moan)  & $-0.49683(7)$  & 44{,}890 \\
  \end{tabular}
  \end{ruledtabular}
  \label{tab:best_seed_energies}
\end{table}

\prsec{Numerical Results}Next, we validate our framework using the square lattice $J_1$-$J_2$ Heisenberg model, whose Hamiltonian is 
\begin{equation}
    \hat{\mathcal{H}}= J_1 \sum_{\langle i,j\rangle} \hat{\mathbf{S}}_{i}\!\cdot\!\hat{\mathbf{S}}_{j}
+ J_2 \sum_{\langle\!\langle i,j\rangle\!\rangle} \hat{\mathbf{S}}_{i}\!\cdot\!\hat{\mathbf{S}}_{j},
\end{equation}
where \(\hat{\mathbf{S}}_{i}=(\hat{S}_{i}^{x},\hat{S}_{i}^{y},\hat{S}_{i}^{z})\) are spin-\(1/2\) operators, \(\langle i,j\rangle\) denotes nearest neighbors, and \(\langle\!\langle i,j\rangle\!\rangle\) next-nearest neighbors.
All numerical experiments are performed on the $10\times 10$ lattice with $J_2/J_1=0.5$ and periodic boundary conditions.
We adopt the standard TQS encoder--decoder architecture~\cite{viteritti2023transformer0,rende2024simple,PhysRevB.111.134411} and perform all calculations within the VMC framework; unless otherwise stated, we report the best energy among five random seeds~\footnote{For all implementation details, including the precise network specification, optimization settings, and computing environment, as well as seed-averaged results and the corresponding standard error of the mean across seeds, see the Supplemental Material.}.

Table~\ref{tab:result1} compares the energies obtained with the PITQS ansatz [Eq.~(\ref{eq:pitqs_U_static})] and the standard TQS ansatz [Eq.~(\ref{eq:time_ordered_product})].
In both cases, we use the first-order Lie--Trotter scheme for each layer.
We fix $\Delta\tau=0.5$ and compare PITQS and TQS at matched total imaginary times $\beta\in\{1.0,2.0,3.0,4.0\}$, i.e., $L\in\{2,4,6,8\}$ layers.
With a constant number of variational parameters $N_p$, the PITQS ansatz achieves energies comparable to the TQS for each $\beta$. For larger $\beta$, the PITQS further achieves lower energies.
In contrast, $N_p$ for the standard TQS grows approximately linearly with $\beta$; nevertheless, increasing $N_p$ by taking larger $\beta$ does not lead to a systematic improvement in accuracy.
These results indicate that conventional TQS with effective Hamiltonians dependent on imaginary time are heavily overparameterized: the essential representational power resides in the structure of the effective Hamiltonian itself.

Furthermore, at a fixed total imaginary time $\beta=2.0$, we benchmark several update strategies to assess how the decomposition schemes affect the accuracy. Table~\ref{tab:best_seed_energies} shows that increasing the decomposition order consistently tends to improve the energy relative to the first-order Lie--Trotter scheme, indicating a systematic reduction of Trotter errors within the sequential family.
When we match the parameter budget to the TQS baseline with $\beta=0.5, N_p=44{,}890$, our PITQS ans\"atze show a substantial performance advantage.
Moreover, compared to standard TQS baselines, the fourth-order Suzuki and Blanes--Moan schemes achieve energies that are better than TQS despite using substantially fewer variational parameters than the deeper TQS model with $\beta=2.0, N_p=155{,}620$. 
However, we note that higher-order decompositions come with trade-offs in computational cost and numerical stability, which can limit their practical advantages depending on the setting~\footnote{For details on the computational cost and numerical stability, see also the Supplemental Material.}.

\begin{figure}[t!]
    \centering
    \includegraphics[width=0.98\linewidth]{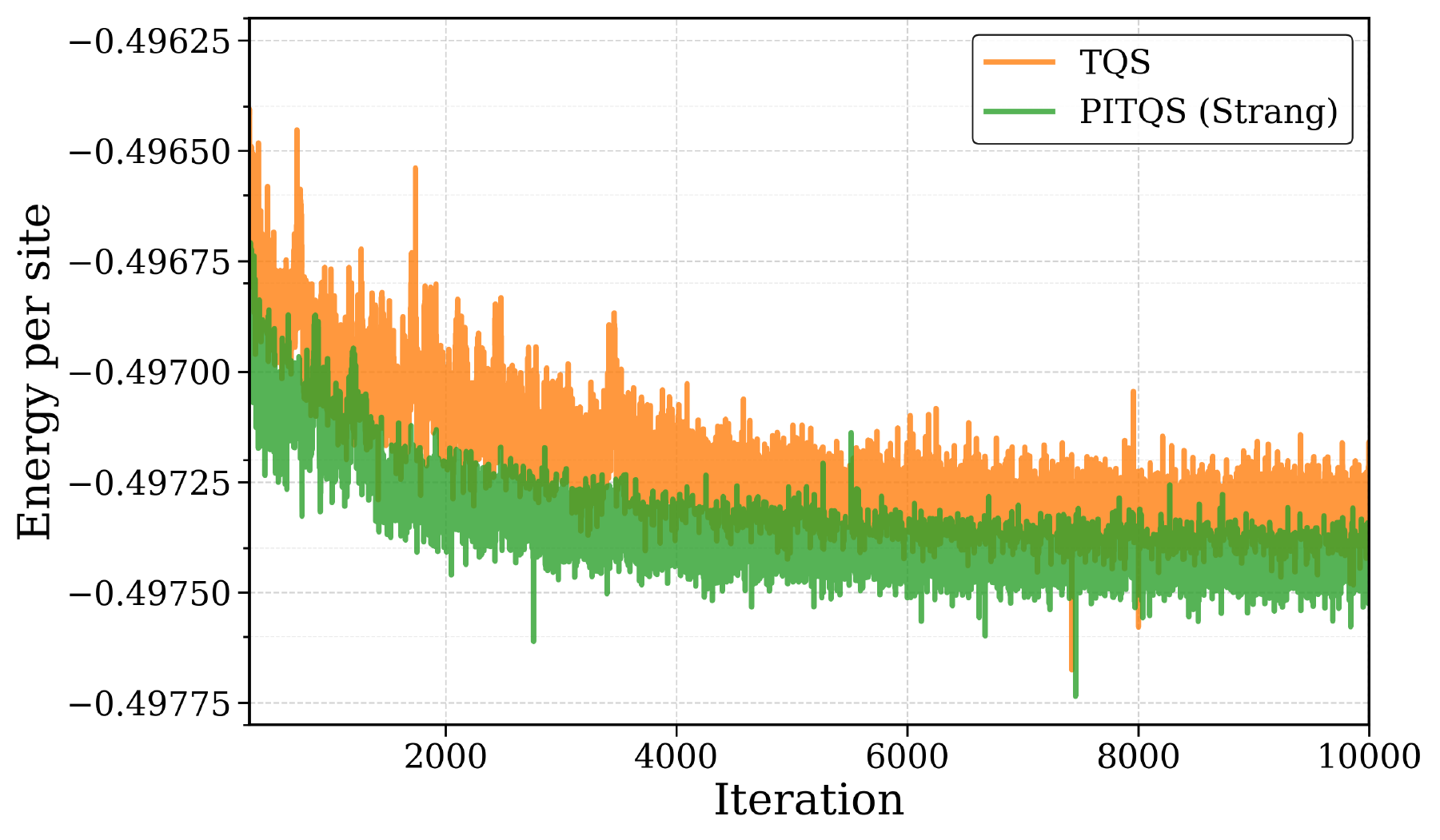}
    \caption{
    Optimization of the energy per site (the first 300 iterations are omitted for readability). We compare the PITQS (Strang) ansatz with $N_p=143{,}010$ against a standard TQS with $N_p=155{,}620$.
    }
    \label{fig:exsota}
\end{figure}

Finally, we demonstrate the scalability of our framework by targeting high accuracy.
We trained the larger PITQS model, a second-order Strang ansatz with $d=108, n_h=18, L=8, \Delta\tau=0.5, N_p=143{,}010$, optimizing it over 10,000 iterations with 8,192 Monte Carlo samples per iteration.
To ensure a fair comparison, we benchmark it against a standard TQS of comparable parameter count ($d=60, L=4, \Delta\tau=1.0, N_p=155{,}620$).
We adopt this baseline because reproducing large-scale TQS~\cite{rende2024simple} ($N_p \approx 268\mathrm{k}$--$995\mathrm{k}$) is computationally prohibitive on our setup.

As illustrated in Fig.~\ref{fig:exsota}, our PITQS ansatz substantially outperforms the comparable standard TQS throughout the optimization process, achieving superior accuracy despite using fewer variational parameters ($N_p=143{,}010$ vs. $155{,}620$).
Most notably, the Strang ansatz achieves an energy of $E \approx -0.49741(3)$, which is competitive with the values reported in Ref.~\cite{rende2024simple} of $E \approx -0.49725$ ($N_p = 267{,}720$) and $E \approx -0.49730$ ($N_p = 994{,}700$), while using only 143,010 parameters.
This highlights that a physics-inspired inductive bias in the ansatz, rather than sheer parameter scale, plays a key role in accuracy.

\prsec{Machine-learning perspective}From a machine-learning perspective, the PITQS ansatz with a single effective Hamiltonian is naturally related to Universal Transformers~\cite{dehghani2018universal}: a single Transformer layer is iterated over depth with \emph{weight sharing}, analogous to recurrence in the broad sense used for recurrent neural networks~\cite{ELMAN1990179}.
Although weight sharing in Transformers is often introduced for parameter efficiency (e.g., ALBERT~\cite{Lan2020ALBERT:}) and can raise concerns about expressivity, in our setting it corresponds to evolution independent of imaginary time and therefore serves as a constructive inductive bias, enhancing physical transparency without sacrificing accuracy and, in some regimes, even enhancing it.

\prsec{Summary}In this Letter, we introduced the physics-inspired framework that reinterprets deep TQS as LITE driven by a single effective Hamiltonian.
By enforcing weight sharing across layers and employing higher-order Trotter--Suzuki decompositions, we showed that PITQS architectures can match state-of-the-art TQS models on the $J_1$-$J_2$ Heisenberg model using substantially fewer variational parameters than standard TQS architectures.
This work bridges the gap between machine-learning expressivity and physical transparency, turning NQS architecture design from heuristic trial-and-error into a systematic and physically grounded construction.

\begin{acknowledgments}
This work was supported by JST CREST Grant Number JPMJCR1913 and JSPS KAKENHI Grant Numbers JP22H00516, JP22H05106 (Y.K.), JP25K15230 (T.K.), and JP24K20864 (I.S.).
\end{acknowledgments}

\section{Supplemental Material}

\section{Implementation of $\hat{\mathcal{K}}_\theta$ and $\hat{\mathcal{V}}_\theta$}

We describe the implementation of the non-local operator $\hat{\mathcal{K}}_\theta$ and the local operator $\hat{\mathcal{V}}_\theta$ used in our numerical experiments for PITQS.
As an illustrative example, we present the layer update in the form of the Lie--Trotter scheme.
A discrete configuration $n$ is first mapped to a latent token state
$z(0)\in\mathbb{R}^{N_{\mathrm{tok}}\times d}$ consisting of $N_{\mathrm{tok}}$ tokens. The $i$-th row $z_i(0) \in \mathbb{R}^d$ is a $d$-dimensional vector for the token at index $i \in \{1,...,N_{\mathrm{tok}}\}$. We then apply a transformation composed of $L$ layers to $z(0)$ to obtain $z(\beta)$.
Following the notation in the main text, we denote the state in the $\ell$-th layer by $z^{(\ell)}\in\mathbb{R}^{N_{\mathrm{tok}}\times d}$. We write the update from $z^{(\ell-1)}$ to $z^{(\ell)}$ as
\begin{align}
\tilde z^{(\ell)}
&= z^{(\ell-1)} + \Delta\tau\,\hat{\mathcal{K}}_\theta\,\!\bigl[z^{(\ell-1)}\bigr],\\
z^{(\ell)}
&= \tilde z^{(\ell)} + \Delta\tau\,\hat{\mathcal{V}}_\theta\,\!\bigl[\tilde z^{(\ell)}\bigr],
\label{eq:preln_absorbed}
\end{align}
where both operators $\hat{\mathcal{K}}_\theta[\cdot]$ and $\hat{\mathcal{V}}_\theta[\cdot]$ internally apply a pre-layer normalization (pre-LN) to their inputs.

The non-local operator $\hat{\mathcal{K}}_\theta$ is implemented by factored MHA (FMHA)~\cite{Rende_2025,rende2024simple}.
FMHA first normalizes the input $z^{(\ell-1)}$ using pre-LN: $\bar z^{(\ell-1)}=\mathrm{LN}(z^{(\ell-1)})$. Then, for each head $h=1,\dots,n_h$, the value features are computed as
\begin{equation}
V^{(h)} = \bar z^{(\ell-1)}\,W_v^{(h)} \in \mathbb{R}^{N_{\mathrm{tok}}\times d_h},
\end{equation}
where $d_h=d/n_h$, and $W_v^{(h)}\in\mathbb{R}^{d\times d_h}$.
Next, FMHA mixes tokens for head $h$ as
\begin{align}
Y^{(h)} = A^{(h)}V^{(h)} \in \mathbb{R}^{N_{\mathrm{tok}}\times d_h}.
\end{align}
Here, $A^{(h)}$ is a trainable, input-independent token--token kernel matrix given by
\begin{align}
A^{(h)}&=[\alpha^{(h)}_{ij}]\in\mathbb{R}^{N_{\mathrm{tok}}\times N_{\mathrm{tok}}},
\end{align}
where $(i,j)$-entry $\alpha^{(h)}_{i,j}$ is the attention weight that represents the effective interaction strength between tokens $i$ and $j$. After concatenating the head outputs $Y^{(h)}$ along the feature dimension and applying an output projection $W_{o}\in\mathbb{R}^{d\times d}$, the map is written as
\begin{equation}
\hat{\mathcal{K}}_\theta[z^{(\ell-1)}]
=
\mathrm{Concat}\,\!\bigl(Y^{(1)},\dots,Y^{(n_h)}\bigr)\,W_o
\in\mathbb{R}^{N_{\mathrm{tok}}\times d}.
\end{equation}
This representation makes the non-locality explicit: the kernel $A^{(h)}$ couples all token pairs $(i,j)$.
For translationally invariant settings, one may further constrain $\alpha^{(h)}_{ij}$ to depend only on relative displacements.

The local operator $\hat{\mathcal{V}}_\theta$ is implemented by a position-wise FFN.
As in $\hat{\mathcal{K}}_\theta$, we absorb pre-LN in the definition of the operator.
Namely, given an input $\tilde z^{(\ell)}$, we first obtain the normalized tokens
$\check z^{(\ell)}=\mathrm{LN}(\tilde z^{(\ell)})$ and then apply the same two-layer MLP independently to each token index $i=1,\dots,N_{\mathrm{tok}}$:
\begin{equation}
\bigl(\hat{\mathcal{V}}_\theta[\tilde z^{(\ell)}]\bigr)_i
=\mathrm{GELU}\!\left(\check z^{(\ell)}_i W_1 + b_1^{\top}\right) W_2 + b_2^{\top}.
\label{eq:ffn}
\end{equation}
Here, $\check z^{(\ell)}_i \in \mathbb{R}^d$ denotes the $i$-th row of $\check z^{(\ell)}$.
$W_1\in\mathbb{R}^{d\times d_{\mathrm{FFN}}}$ and $W_2\in\mathbb{R}^{d_{\mathrm{FFN}}\times d}$ are weight matrices,
$b_1\in\mathbb{R}^{d_{\mathrm{FFN}}}$ and $b_2\in\mathbb{R}^{d}$ are bias vectors, and the hidden dimension is chosen as $d_{\mathrm{FFN}}=4d$.
This map is explicitly local (on-site): the output at index $i$ depends only on the specific input token $\tilde z^{(\ell)}_i$ and is not mixed with the other tokens.

\section{PITQS for Fermion systems}
\label{sec:FPITQS}

We introduce PITQS tailored to fermionic many-body systems.
Compared with the spin-system PITQS, there are three essential differences: (i) the encoder that maps an occupation configuration to tokens, (ii) the implementation of the LITE block, and (iii) the incorporation of antisymmetry in the decoder.
Following the construction of fermionic TQS in, e.g., Refs.~\cite{qxc3-bkc7,gu2025solvinghubbardmodelneural}, we build fermionic PITQS as follows.

We represent a fermionic occupation configuration by $n\in\{0,1\}^{2N}$.
For each spatial site $i=1,\dots,N$, we consider a local class label $c_i$ that takes one of four values and encodes the on-site occupancy:
\begin{equation}
c_i \equiv n_{i,\uparrow} + 2\,n_{i,\downarrow} \in \{0,1,2,3\},
\label{eq:occ_class}
\end{equation}
where $n_{i\sigma}$ denotes the occupation number of spin $\sigma$ at site $i$.
The encoder embeds these discrete labels with a learned continuous embedding matrix $E\in\mathbb{R}^{4\times d}$, producing token features
\begin{equation}
z_i(0) = E_{c_i} + p_i \in \mathbb{R}^{d},
\qquad i=1,\dots,N,
\label{eq:fermion_tok}
\end{equation}
where $\{p_i\}_{i=1}^{N}$ are learned positional embeddings. Collecting tokens gives
$z(0)\in\mathbb{R}^{N\times d}$, and hence we set $N_{\mathrm{tok}}=N$ in the fermionic setting.

Next, the non-local operator $\hat{\mathcal{K}}_\theta$ is implemented by standard MHA~\cite{NIPS2017_3f5ee243} with queries and keys, following standard fermionic TQS architectures~\cite{qxc3-bkc7,gu2025solvinghubbardmodelneural}.
As in FMHA, we first normalize the input $z^{(\ell-1)}\in\mathbb{R}^{N_{\mathrm{tok}}\times d}$ as $\bar z^{(\ell-1)}=\mathrm{LN}(z^{(\ell-1)})$. Then, for each head $h=1,\dots,n_h$ with $d_h=d/n_h$, we compute the query, key, and value matrices,
\begin{align}
Q^{(h)} &= \bar z^{(\ell-1)} W_Q^{(h)},\\
K^{(h)} &= \bar z^{(\ell-1)} W_K^{(h)},\\
V^{(h)} &= \bar z^{(\ell-1)} W_V^{(h)},
\label{eq:mhsa_qkv}
\end{align}
where $W_Q^{(h)},W_K^{(h)},W_V^{(h)}\in\mathbb{R}^{d\times d_h}$.
The (input-dependent) token--token kernel matrix is defined by the row-wise softmax function
\begin{equation}
A^{(h)}(\bar z^{(\ell-1)})=\mathrm{Softmax}\!\left(\frac{Q^{(h)}K^{(h)\top}}{\sqrt{d_h}}\right).
\label{eq:mhsa_kernel}
\end{equation}
The head output is then written in the same form as FMHA,
\begin{equation}
Y^{(h)} = A^{(h)}(\bar z^{(\ell-1)})\,V^{(h)} \in \mathbb{R}^{N_{\mathrm{tok}}\times d_h},
\label{eq:mhsa_headout}
\end{equation}
and after concatenation and an output projection $W_{o}\in\mathbb{R}^{d\times d}$, we obtain
\begin{align}
\hat{\mathcal{K}}_\theta[z^{(\ell-1)}]
=
\mathrm{Concat}\,\!\bigl(Y^{(1)},\dots,Y^{(n_h)}\bigr)\,W_o
\in\mathbb{R}^{N_{\mathrm{tok}}\times d}.
\label{eq:mhsa_map}
\end{align}
This representation makes the non-locality explicit: $A^{(h)}(z^{(\ell-1)})$ couples all token pairs $(i,j)$ and is state-dependent through $Q^{(h)}$ and $K^{(h)}$.
The local operator $\hat{\mathcal{V}}_\theta$ follows the same structure as Eq.~(\ref{eq:ffn}), absorbing the pre-LN step.
Here, we employ SiLU as the nonlinearity instead of GELU and set the hidden dimension to $d_{\mathrm{FFN}}=2d$:
\begin{equation}
\bigl(\hat{\mathcal{V}}_\theta[\tilde z^{(\ell)}]\bigr)_i
=\mathrm{SiLU}\!\left(\check z^{(\ell)}_i W_1 + b_1^{\top}\right) W_2 + b_2^{\top},
\label{eq:ffn_silu}
\end{equation}
where $\check z^{(\ell)}_i$ denotes the $i$-th row of the normalized tokens $\mathrm{LN}(\tilde z^{(\ell)})$.
The parameter shapes are identical to those in Eq.~(\ref{eq:ffn}) except for the hidden dimension.
Finally, we enforce weight sharing across layers, i.e., the PITQS has a single static effective Hamiltonian
\begin{equation}
\hat{\mathcal{H}}_\theta = -(\hat{\mathcal{V}}_\theta + \hat{\mathcal{K}}_\theta),
\end{equation}
and the LITE operator is $\hat{\mathcal{U}}_\theta(\beta) = e^{-\beta \hat{\mathcal{H}}_\theta}$.
In practice, $\hat{\mathcal{U}}_\theta(\beta)$ is realized by applying a shared Transformer layer with $\hat{\mathcal{V}}_\theta$ and $\hat{\mathcal{K}}_\theta$ $L$ times. The layer is implemented as a Trotter--Suzuki decomposition of the exponential (with step size $\Delta\tau=\beta/L$).

For the decoder $\hat{\mathcal{D}}_\theta$, we employ a multi-Slater backflow~\cite{qxc3-bkc7,gu2025solvinghubbardmodelneural} to enforce antisymmetry.
From each decoded token $z_{i}(\beta)$, we generate configuration-dependent backflow orbitals by a shared, position-wise affine map,
\begin{equation}
  x_i(\beta)=\mathrm{LN}(z_i(\beta))\,W_{\mathrm{out}}+\mathbf{1}b_{\mathrm{out}}^\top,
\end{equation}
with $W_{\mathrm{out}}\in\mathbb{R}^{d\times (2KN_e)}$ and $b_{\mathrm{out}}\in\mathbb{R}^{2KN_e}$.
The vector $x_i$ is interpreted as $K$ sets of spin-resolved orbitals: for each determinant $k \in \{1,...,K\}$, we extract two
$N_e$-dimensional vectors $x_{i,\downarrow}^{(k)}$ and $x_{i,\uparrow}^{(k)}$, which we regard as the
rows of a backflow orbital matrix $M^{(k)}(n)\in\mathbb{R}^{2N\times N_e}$.
Concretely, the row of $M^{(k)}(n)$ associated with the spin-orbital $(i,\sigma)$ is $x_{i,\sigma}^{(k)}$ in the
chosen spin-orbital ordering.
Given an occupation configuration $n\in\{0,1\}^{2N}$, let $P(n)\in\{0,1\}^{N_e\times 2N}$ be a row-selection matrix
that picks the occupied spin orbitals. Then, the $k$-th Slater matrix $\Phi^{(k)}_{\theta}(n)$ is represented as a matrix product:
\begin{equation}
  \Phi^{(k)}_{\theta}(n) = P(n)\,M^{(k)}_{\theta}(n)\in\mathbb{R}^{N_e\times N_e}.
\end{equation}
Finally, the decoder outputs the complex log-amplitude corresponding to a real multi-determinant backflow form,
\begin{align}
\hat{\mathcal{D}}_\theta[z(\beta)]
&=
\log\left|\sum_{k=1}^{K}\det \Phi^{(k)}_{\theta}(n)\right|\\
&+ i\,\pi\,\mathbb{I}\!\left[\sum_{k=1}^{K}\det \Phi^{(k)}_{\theta}(n)<0\right],
\end{align}
where $\mathbb{I}\,[\cdot]$ is the indicator function.

\section{Experimental Details}
\label{sec:expdetails}

All numerical experiments were carried out in the VMC framework, where the variational energy is estimated from Monte Carlo samples $n\sim p_\theta(n)$ with $p_\theta(n)\propto|\Psi_\theta(n)|^{2}$ as $\mathbb{E}_{n\sim p_\theta}\,\![(\hat{\mathcal H}\Psi_\theta)(n)/\Psi_\theta(n)]$.
Our PITQS architectures were implemented in the NetKet library~\cite{netket2:2019,netket3:2022}, which is built on JAX~\cite{jax2018github}, Flax~\cite{flax2020github}, and mpi4jax~\cite{Häfner2021}.
Unless otherwise stated, we report the best final energy obtained after 800 iterations among five independent runs with random initializations.
For a detailed analysis of stability across random seeds, see
Sec.~\hyperref[sec:stability]{COMPUTATIONAL STABILITY OF PITQS} in this Supplemental Material.
Computations were performed on two NVIDIA A100 GPUs with 40~GB of memory each for models that fit in memory, while larger models were run on two NVIDIA H100 GPUs with 80~GB each.

For the $J_1$-$J_2$ Heisenberg model, the variational parameters were optimized using the minimum-step stochastic reconfiguration (MinSR)~\cite{chen2024empowering,rende2024simple} with a learning rate of $0.0075$ and a diagonal shift of $10^{-4}$.
Unless otherwise specified, we performed 800 optimization iterations with 4,096 Monte Carlo samples per iteration, and set the hyperparameters to model dimension $d=60$, patch size $2\times2$, number of heads $n_h=10$, and FFN hidden dimension $4d$.
The optimization budget and these hyperparameters follow the baseline setting of the NetKet tutorial on the Vision Transformer wave function: see~\url{https://netket.readthedocs.io/en/latest/tutorials/ViT-wave-function.html}.

For the Hubbard model, we optimized the variational parameters using MinSR with a learning rate of $0.01$ and a diagonal shift of $10^{-4}$.
We performed $10{,}000$ optimization iterations with $4{,}096$ Monte Carlo samples per iteration.
The hyperparameters were set to model dimension $d=16$, number of heads $n_h=4$, depth $L=4$, and imaginary-time step $\Delta\tau=0.5$.

\section{Computational stability of PITQS}
\label{sec:stability}

\begin{table}[t!]
  \centering
  \caption{
    Energies per site averaged over five random seeds, the corresponding SEM across seeds, and the number of variational parameters $N_p$ for the PITQS and TQS ans\"atze at $\beta=1.0,2.0,3.0,4.0$.
  }
  \begin{ruledtabular}
    \begin{tabular}{l c c c}
      \multicolumn{4}{c}{$\beta=1.0$} \\
      Scheme & Mean & SEM ($\times10^{-4}$) & $N_p$  \\
      \hline
      TQS & $-0.49576$ & $0.82$ & $81{,}800$ \\
      PITQS (Lie--Trotter) & $-0.49579$ & $0.56$ & $44{,}890$ \\
      PITQS (Strang) & $-0.49563$ & $2.3$ & $44{,}890$ \\
      PITQS (Suzuki) & $-0.49642$ & $0.88$ & $44{,}890$ \\
      PITQS (Blanes--Moan) & $-0.49578$ & $3.6$ & $44{,}890$ \\
      \hline

      \multicolumn{4}{c}{$\beta=2.0$} \\
      Scheme & Mean & SEM ($\times10^{-4}$) & $N_p$ \\
      \hline
      TQS & $-0.49583$ & $4.8$ & $155{,}620$ \\
      PITQS (Lie--Trotter) & $-0.49626$ & $1.2$ & $44{,}890$ \\
      PITQS (Strang) & $-0.49633$ & $1.3$ & $44{,}890$ \\
      PITQS (Suzuki) & $-0.49653$ & $2.7$ & $44{,}890$ \\
      PITQS (Blanes--Moan) & $-0.49672$ & $0.32$ & $44{,}890$ \\
      \hline

      \multicolumn{4}{c}{$\beta=3.0$} \\
      Scheme & Mean & SEM ($\times10^{-4}$) & $N_p$ \\
      \hline
      TQS & $-0.49621$ & $2.0$ & $229{,}440$ \\
      PITQS (Lie--Trotter) & $-0.49669$ & $0.34$ & $44{,}890$ \\
      PITQS (Strang) & $-0.49665$ & $0.35$ & $44{,}890$ \\
      PITQS (Suzuki) & $-0.49608$ & $2.9$ & $44{,}890$ \\
      PITQS (Blanes--Moan) & $-0.49658$ & $1.1$ & $44{,}890$ \\
      \hline

      \multicolumn{4}{c}{$\beta=4.0$} \\
      Scheme & Mean & SEM ($\times10^{-4}$) & $N_p$ \\
      \hline
      TQS & $-0.49645$ & $0.27$ & $303{,}260$ \\
      PITQS (Lie--Trotter) & $-0.49654$ & $2.5$ & $44{,}890$ \\
      PITQS (Strang) & $-0.49659$ & $0.64$ & $44{,}890$ \\
      PITQS (Suzuki) & $-0.49630$ & $2.8$ & $44{,}890$ \\
      PITQS (Blanes--Moan) & $-0.49613$ & $3.6$ & $44{,}890$ \\
    \end{tabular}
  \end{ruledtabular}
  \label{tab:stability}
\end{table}

We examine how the computational stability of PITQS depends on the total imaginary time $\beta=L\Delta\tau$. 
While the main text reports the best energy obtained among independent runs, here we focus on the run-to-run variability and evaluate the summary statistics, namely the mean and standard error.
We set the step size to $\Delta\tau=0.5$ and sweep $\beta$ over $\{1.0, 2.0, 3.0, 4.0\}$ while keeping all other settings fixed. These values of $\beta$ correspond to the depths $L \in \{2, 4, 6, 8\}$, respectively.
To assess stability, we repeat each experiment with five independent random seeds and summarize the resulting run-to-run variability.
The experiment is performed on two NVIDIA A100 GPUs, except for the standard TQS baseline at $\beta \in \{3.0, 4.0\}$, which is run on two NVIDIA H100 GPUs due to memory constraints.

Table~\ref{tab:stability} reports the energy per site averaged over five random seeds together with the corresponding standard error of the mean (SEM) across seeds.
In the small-$\beta$ regime ($\beta=1.0, 2.0$), higher-order decompositions (Suzuki and Blanes--Moan) exhibit superior accuracy, yielding lower mean energies.
However, in the large-$\beta$ regime ($\beta=3.0, 4.0$), the performance of these high-order methods deteriorates significantly; they suffer from both higher mean energies and increased SEMs, indicating optimization instability.
Conversely, the low-order Lie--Trotter and Strang schemes maintain high robustness even at larger $\beta$, and are more stable than the higher-order schemes while achieving competitive or better energies than TQS.
These observations highlight a practical trade-off for higher-order decompositions: while they can reduce Trotter error at moderate $\beta$, their increased optimization sensitivity at larger $\beta$ warrants care when selecting $(L,\Delta\tau)$ in practice.

\section{Computational cost of PITQS}

We evaluate the computational cost of PITQS and the standard TQS in terms of two practical metrics: the total peak usage of GPU memory and the computational time per optimization iteration.
All benchmarks are performed at fixed step size $\Delta\tau=0.5$ for total imaginary times $\beta\in\{1.0,2.0,3.0,4.0\}$, and all computational costs are measured in separate profiling runs on two NVIDIA H100 GPUs using identical hyperparameters.

\begin{figure}[t!]
    \centering
    \includegraphics[width=0.9\linewidth]{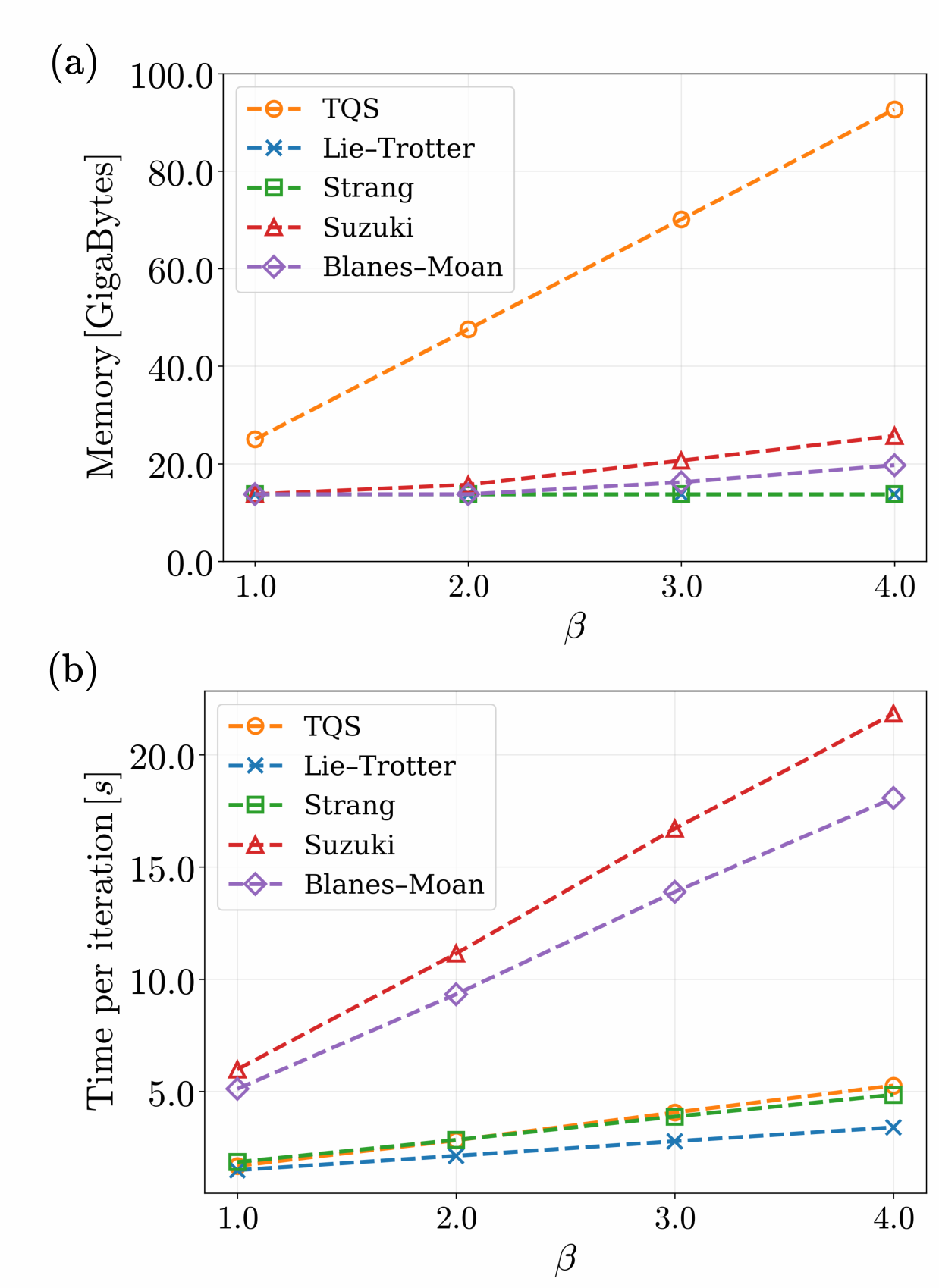}
    \caption{
    Computational costs of TQS and PITQS as a function of total imaginary time $\beta\in\{1.0,2.0,3.0,4.0\}$. (a) Peak GPU memory usage. (b) Computational time per optimization iteration in seconds. 
    }
    \label{fig:result_SI_cost}
\end{figure}

Fig.~\ref{fig:result_SI_cost}~(a) compares the peak GPU memory usage of PITQS and the standard TQS.
For PITQS, the low-order Lie--Trotter and Strang schemes exhibit an essentially $\beta$-independent memory usage, consistent with weight sharing in the LITE blocks, which keeps the number of variational parameters independent of the unrolling depth.
Even the fourth-order Suzuki and Blanes--Moan decompositions show only a moderate increase at larger $\beta$, owing to the intermediate buffers required for higher-order stages.
In contrast, the memory usage of TQS increases strongly with $\beta$, reflecting that increasing $\beta$ requires increasing the depth $L=\beta/\Delta\tau$ and thus the number of distinct layer parameters.
Overall, PITQS remains substantially more memory-efficient than TQS at larger $\beta$.

Fig.~\ref{fig:result_SI_cost}~(b) shows the computational time per iteration for optimizing TQS and PITQS.
We find that the computational times of the Lie--Trotter and Strang schemes of PITQS are comparable to those of TQS, although TQS exhibits a mild upward trend at larger $\beta$, consistent with its substantially larger parameter count relative to PITQS.
In contrast, the fourth-order Suzuki and Blanes--Moan decompositions incur a significantly higher cost, typically by a factor of $\sim5$--$7$ compared with the lower-order schemes.
This increase reflects the larger number of stages $k$ in fourth-order decompositions and highlights a practical trade-off between accuracy and computational cost for higher-order schemes.
Taken together, these results suggest that the Lie--Trotter or Strang schemes provide a favorable balance of computational cost, memory efficiency, and accuracy in many practical settings.

\section{Cooling profile in latent space}

To directly probe whether the LITE acts as a cooling procedure, we evaluate the energy of a pre-trained PITQS at intermediate imaginary-time extents.
Specifically, we take the PITQS ansatz with the Strang scheme used in Fig.~2 of the main text and evaluate the energy by truncating the unrolling length to $\beta=0.5,\dots,4.0$ while keeping the variational parameters $\theta^*$ pre-trained at $\beta_{\mathrm{train}}=4.0$ fixed.
Following the main text, we define the imaginary-time extent as $\beta=L\Delta\tau$.
In the present benchmark, we fix $\Delta\tau=0.5$, which yields eight evaluation points $\beta\in\{0.5,1.0,1.5,2.0,2.5,3.0,3.5,4.0\}$.
This procedure defines a family of wave functions $\Psi^{(\beta)}_{\theta^{*}}$ obtained by applying $L=\beta/\Delta\tau$ shared evolution blocks followed by the same decoder $\hat{\mathcal{D}}_{\theta^{*}}$.

For each $\beta$, we estimate the energy per site $E^{(\beta)}$ with fixed trained variational parameters $\theta^{*}$.
In particular, we use $8{,}192$ Monte Carlo samples for the evaluation at each $\beta$.

\begin{figure}[t!]
    \centering
    \includegraphics[width=0.95\linewidth]{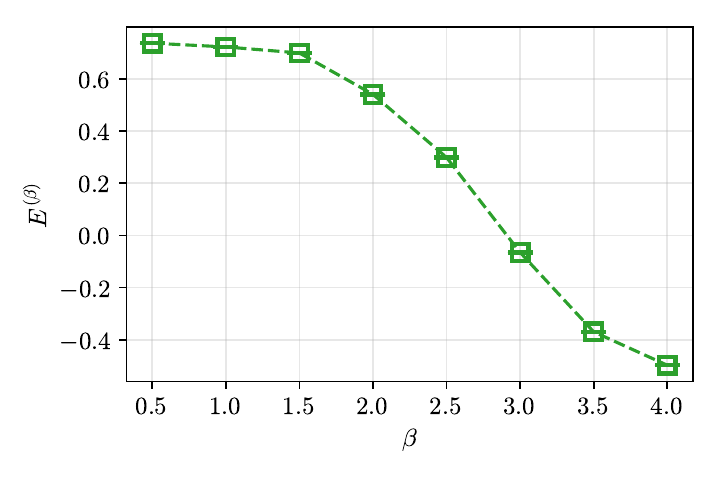}
    \caption{Energies per site $E^{(\beta)}$ evaluated using $8{,}192$ Monte Carlo samples by truncating the PITQS ansatz using the Strang scheme trained at $\beta=4.0$. Error bars show the Monte Carlo error.}
    \label{fig:cooling}
\end{figure}

Fig.~\ref{fig:cooling} shows the resulting estimates of the energy per site $E^{(\beta)}$ for eight evaluation points.
Within the training horizon, increasing $\beta$ systematically lowers the variational energy.
This behavior indicates that iterating the LITE steps in the latent space acts as an effective cooling procedure, consistent with the LITE interpretation.

The evaluation is restricted to $\beta\le\beta_{\mathrm{train}}=8\Delta\tau$, that is, within the training horizon, and does not claim reliable extrapolation beyond it.
In practice, evaluating at $\beta>\beta_{\mathrm{train}}$ corresponds to extrapolating the latent evolution beyond the training horizon, where the decoder may experience a distribution shift.
Therefore, obtaining reliable predictions at larger $\beta$ requires fine-tuning with the extended unrolling length.

\section{Dependence on the imaginary-time step size $\Delta\tau$}

We examine how the energy per site depends on the imaginary-time step size $\Delta\tau$ used to discretize the LITE operator.
We treat $\Delta\tau$ as a hyperparameter and evaluate the three values $\Delta\tau \in \{0.1, 0.5, 1.0\}$ with the depth fixed at $L=4$, i.e., $\beta\in\{0.4,2.0,4.0\}$.

\begin{table*}[t!]
  \centering
  \caption{
    Energy per site for $\Delta\tau \in \{0.1, 0.5, 1.0\}$ at fixed depth $L=4$. Parentheses indicate Monte Carlo errors, and $N_p$ denotes the number of variational parameters.
  }
  \begin{ruledtabular}
    \begin{tabular}{lcccc}
      \multicolumn{1}{c}{Scheme}
      & \multicolumn{1}{c}{$\Delta\tau=0.1$}
      & \multicolumn{1}{c}{$\Delta\tau=0.5$}
      & \multicolumn{1}{c}{$\Delta\tau=1.0$}
      & \multicolumn{1}{c}{$N_p$} \\
      \hline
      TQS
      & $-0.49620(11)$ & $-0.49652(9)$ & $-0.49666(9)$ & $155{,}620$ \\
      PITQS (Lie--Trotter)
      & $-0.49681(10)$ & $-0.49669(9)$ & $-0.49659(10)$ & $44{,}890$ \\
      PITQS (Strang)
      & $-0.49667(9)$ & $-0.49671(9)$ & $-0.49667(9)$ & $44{,}890$ \\
      PITQS (Suzuki)
      & $-0.49678(8)$ & $-0.49697(9)$ & $-0.49696(9)$ & $44{,}890$ \\
      PITQS (Blanes--Moan)
      & $-0.49679(10)$ & $-0.49683(7)$ & $-0.49701(11)$ & $44{,}890$ \\
    \end{tabular}
  \end{ruledtabular}
  \label{tab:dtau_dependence}
\end{table*}

Table~\ref{tab:dtau_dependence} summarizes the resulting energies per site and lists $N_p$ for reference.
The column at $\Delta\tau=0.5$ corresponds to the setting used in the main text.
At fixed depth $L=4$, the dependence on $\Delta\tau$ varies across the schemes, reflecting their distinct theoretical properties.
For PITQS, the observed trends are consistent with the theoretical properties of the Trotter--Suzuki decompositions.
The Lie--Trotter scheme yields slightly higher energies as $\Delta\tau$ increases, which is expected for a first-order approximation.
The Strang decomposition shows no significant degradation within the statistical uncertainty, consistent with its second-order nature.
Notably, the fourth-order Suzuki and Blanes--Moan decompositions maintain high accuracy even at $\Delta\tau=1.0$, demonstrating the benefit of higher-order decompositions at larger $\Delta\tau$.

TQS exhibits a trend opposite to PITQS with the Lie--Trotter scheme. Although TQS effectively implements the first-order Lie--Trotter decomposition, we confirm that its accuracy improves with larger $\Delta\tau$.
Since the depth $L$ is fixed, increasing $\Delta\tau$ implies a larger total imaginary time $\beta$, which generally aids ground-state cooling.
This behavior of TQS is likely attributable to its layer-dependent effective Hamiltonians, which lead to a larger variational freedom and allow TQS to absorb finite-$\Delta\tau$ Trotter errors.

\section{Continuous limit and Runge--Kutta integration}

Our \emph{operator-centric} LITE formulation admits a continuous latent imaginary-time limit, echoing the viewpoint of neural ordinary differential equations~\cite{10.5555/3327757.3327764}: the effective Hamiltonian defines a latent-space dynamics whose solution is approximated by a depth-discretized update. This discretization introduces two distinct sources of approximation error.

First, when the effective Hamiltonian is decomposed into noncommuting components (i.e., $\hat{\mathcal{K}}_\theta$ and $\hat{\mathcal{V}}_\theta$), Trotter--Suzuki decompositions incur a Trotter error, which we can reduce by using higher-order decompositions. Second, each component causes numerical integration errors when we approximate its action with discrete updates. To mitigate this error, we can adopt higher-order numerical methods instead of the first-order Euler method presented in ``Latent imaginary-time evolution'' and used elsewhere in our experiments. Specifically, we implement a second-order Runge--Kutta (RK2) method, i.e., Heun's method~\cite{heun1900neue}. For each component $\hat{\mathcal{A}}_\theta \in \{\hat{\mathcal{K}}_\theta, \hat{\mathcal{V}}_\theta\}$, this method updates the latent state from $z$ to $z'$ using the following predictor--corrector step:
\begin{equation}
\begin{aligned}
  k_1 &= \hat{\mathcal{A}}_\theta[z], \\
  \tilde{z} &= z + \Delta\tau k_1, \\
  z' &= z + \frac{\Delta\tau}{2} \left( k_1 + \hat{\mathcal{A}}_\theta[\tilde{z}] \right).
\end{aligned}
\end{equation}
This approach reduces the local integration error, providing a more accurate approximation of the exponential operators $e^{\Delta\tau\hat{\mathcal{K}}_\theta}$ and $e^{\Delta\tau\hat{\mathcal{V}}_\theta}$.

\begin{table}[t!]
  \centering
    \caption{
    Energies per site $E$ obtained with the Euler and the RK2 sub-steps at identical total imaginary times.
  }
  \begin{ruledtabular}
    \begin{tabular}{ccc}
      $\beta$ & $E\,(\text{Euler})$ & $E\,(\text{RK2})$ \\
      \hline
      1.0 & $-0.49596(14)$ & $-0.49628(11)$ \\
      2.0 & $-0.49669(9)$ & $-0.49682(10)$  \\
      3.0 & $-0.49676(10)$ & $-0.49679(10)$  \\
      4.0 & $-0.49696(9)$ & $-0.49695(8)$  \\
    \end{tabular}
  \label{tab:rk2_results}
  \end{ruledtabular}
\end{table}

Table~\ref{tab:rk2_results} reports the energies per site of the Euler and RK2 methods with the Lie--Trotter scheme. The results of the Euler method correspond to the ones of PITQS in Table~I of the main text.
At $\beta=1.0$ and $\beta=2.0$, the RK2 method yields a modest improvement over the Euler method within statistical uncertainty, whereas at $\beta=3.0$ and $ \beta=4.0$ the two methods become statistically indistinguishable.
These results suggest that at larger $\beta$, the potential accuracy gain from RK2 is masked by optimization instability, mirroring the behavior observed in the high-order Suzuki and Blanes--Moan decompositions.

\section{Parallel formulation}

Throughout the main text and in the preceding sections of this Supplemental Material, we primarily employ a \textit{serialized} formulation in which the non-local and local components are applied sequentially within each layer.
For the first-order Lie--Trotter scheme, a single step with $\Delta\tau=\beta/L$ can be written as
\begin{align}
\tilde z^{(\ell)}
&= z^{(\ell-1)} + \Delta\tau\,\hat{\mathcal{K}}_\theta\,\!\bigl[z^{(\ell-1)}\bigr], \\
z^{(\ell)}
&= \tilde z^{(\ell)} + \Delta\tau\,\hat{\mathcal{V}}_\theta\,\!\bigl[\tilde z^{(\ell)}\bigr].
\label{eq:serial}
\end{align}
Equivalently, using $\hat{\mathcal{H}}_\theta = -(\hat{\mathcal{V}}_\theta+\hat{\mathcal{K}}_\theta)$, this serialized formulation corresponds to approximating the full evolution operator
$\hat{\mathcal{U}}_\theta(\Delta\tau)=e^{-\Delta\tau\hat{\mathcal{H}}_\theta}=e^{\Delta\tau(\hat{\mathcal{K}}_\theta+\hat{\mathcal{V}}_\theta)}$
by a product of exponentials of the components.
When $\hat{\mathcal{K}}_\theta$ and $\hat{\mathcal{V}}_\theta$ do not commute, this decomposition incurs a Trotter error, which motivates the use of higher-order decompositions discussed elsewhere in this work.

Here we consider an alternative \textit{parallel} formulation, where the effective Hamiltonian is kept in the exact sum form $\hat{\mathcal{H}}_\theta = -(\hat{\mathcal{V}}_\theta + \hat{\mathcal{K}}_\theta)$ and the update is performed without operator decomposition.
In this formulation, the update with the Euler method follows
\begin{equation}
  z^{(\ell)} = e^{-\Delta\tau \hat{\mathcal{H}}_\theta}[z^{(\ell-1)}] \approx z^{(\ell-1)} - \Delta\tau\hat{\mathcal{H}}_\theta\,[z^{(\ell-1)}].
  \label{eq:parallel}
\end{equation}
Because the formulation uses the full effective Hamiltonian $\hat{\mathcal{H}}_\theta$ directly, it does not introduce a Trotter error associated with splitting $\hat{\mathcal{K}}_\theta$ and $\hat{\mathcal{V}}_\theta$.
In recent machine learning, related parallel formulations have been adopted in large-scale models to reduce serialized computation within each layer~\cite{JMLR:v24:22-1144,pmlr-v202-dehghani23a}.
However, in practice, the parallel structure can be more sensitive during optimization and may require stabilization.

\begin{table}[t!]
  \centering
  \caption{
    Energies per site $E$ obtained with the parallel and serialized formulations at identical total imaginary times.
  }
  \begin{ruledtabular}
    \begin{tabular}{ccccc}
      $\beta$ & $E\,(\text{Serialized})$ & $N_p\,(\text{Serialized})$ & $E\,(\text{Parallel})$ & $N_p\,(\text{Parallel})$ \\
      \hline
      1.0 & $-0.49596(14)$ & $44{,}890$ & $-0.49285(16)$ & $44{,}770$ \\
      2.0 & $-0.49669(9)$  & $44{,}890$ & $-0.49589(11)$ & $44{,}770$ \\
      3.0 & $-0.49676(10)$  & $44{,}890$ & $-0.49647(11)$ & $44{,}770$ \\
      4.0 & $-0.49696(9)$  & $44{,}890$ & $-0.49665(10)$ & $44{,}770$ \\
    \end{tabular}
  \end{ruledtabular}
  \label{tab:parallel_results}
\end{table}

Table~\ref{tab:parallel_results} compares the results of the parallel and serialized formulations under identical hyperparameters at fixed step size $\Delta\tau=0.5$ and $\beta\in\{1.0,2.0,3.0,4.0\}$. The serialized results correspond to the ones of PITQS in Table~I of the main text.
Across the tested range, the serialized formulation achieves slightly lower variational energies than the parallel scheme, while the parameter counts remain nearly identical.
The performance gap is most pronounced at $\beta=1.0$ and becomes smaller at larger $\beta$.
This trend suggests that, for the present model sizes and optimization settings, the practical advantage of the serialized update is not limited by Trotter error alone; rather, the sequential application of $\hat{\mathcal{K}}_\theta$ and $\hat{\mathcal{V}}_\theta$ within each layer can provide a more favorable inductive bias and optimization behavior than the parallel coupling in Eq.~(\ref{eq:parallel}).

\section{Benchmark for Hubbard models}

\begin{table}[t!]
  \centering
  \caption{
    Energy per site $E$ for the $4\times4$ Hubbard model at half filling for $U/t=4$ and $8$  with periodic boundary conditions.
    $N_p$ denotes the number of variational parameters.
    Reference values from exact diagonalization (ED) reported in Ref.~\cite{ANDERSON201322} are also shown.
  }
  \begin{ruledtabular}
  \begin{tabular}{lccc}
    Method & $E\,(U/t=4)$ & $E\,(U/t=8)$ & $N_{p}$ \\
    \hline
    ED & $-0.85137$ & $-0.52931$ & N/A \\
    TQS & $-0.85114(13)$ & $-0.52909(25)$  &  $11{,}424$  \\
    PITQS  & $-0.85113(22)$ & $-0.52919(35)$  & $4{,}752$  \\
  \end{tabular}
  \end{ruledtabular}
  \label{tab:hubbard}
\end{table}

We benchmark PITQS on the two-dimensional Hubbard model and compare with the standard fermionic TQS.
The Hubbard Hamiltonian is given by
\begin{equation}
\hat{\mathcal{H}} = -t \sum_{\langle i,j \rangle, \sigma} \left( \hat{c}_{i\sigma}^{\dagger} \hat{c}_{j\sigma} + \text{H.c.}\right) + U \sum_{i} \hat{n}_{i\uparrow} \hat{n}_{i\downarrow},
\label{eq:hubbard}
\end{equation}
where $\hat{c}_{i\sigma}^{\dagger}$ ($\hat{c}_{i\sigma}$) creates (annihilates) an electron with spin $\sigma \in \{\uparrow, \downarrow\}$ at site $i$, and $\hat{n}_{i\sigma} = \hat{c}_{i\sigma}^{\dagger} \hat{c}_{i\sigma}$ is the number operator.
The first term describes the hopping between nearest-neighbor sites $\langle i, j \rangle$ with amplitude $t$, and the second term represents the on-site Coulomb repulsion with strength $U$.
The implementation and experimental details follow
Sec.~\hyperref[sec:FPITQS]{PITQS FOR FERMION SYSTEMS}
and Sec.~\hyperref[sec:expdetails]{EXPERIMENTAL DETAILS}
in this Supplemental Material.

Table~\ref{tab:hubbard} summarizes the energies for the $4\times 4$ Hubbard model at half filling with periodic boundary conditions, comparing PITQS and TQS at interaction strengths $U/t=4$ and $U/t=8$ under identical hyperparameters.
As in the $J_1$-$J_2$ Heisenberg benchmark in Table~I of the main text, PITQS achieves an accuracy comparable to that of TQS while using substantially fewer variational parameters.
These results indicate that enforcing a static effective Hamiltonian through weight sharing provides a parameter-efficient inductive bias that remains effective beyond spin systems, including interacting fermions.

\bibliography{apssamp}

\end{document}